\RequirePackage{etoolbox}
\csdef{input@path}{{style/}{graphics/}}
\documentclass[MSNbibl,nameyear,dvips]{arxstspdf}
\usepackage{graphicx}
\usepackage{url,breakurl}
\usepackage{flushend}
\usepackage{stfloats}

\volume{30}
\issue{1}
\pubyear{2015}
\firstpage{133}
\lastpage{146}
\doi{10.1214/14-STS495} 
\docsubty{FLA}


\begin{document}
\begin{frontmatter}

\title{A Conversation with Nancy Flournoy}
\runtitle{A Conversation with Nancy Flournoy}

\begin{aug}
\author[A]{\fnms{William F.}~\snm{Rosenberger}\corref{}\ead[label=e1]{wrosenbe@gmu.edu}}
\runauthor{W. F. Rosenberger}

\affiliation{George Mason University}

\address[A]{William F. Rosenberger is University Professor and Chairman,
Department of Statistics,
George Mason University, 4400 University Drive MS 4A7, Fairfax,
Virginia 22030-4444, USA \printead{e1}.}
\end{aug}

%
\begin{abstract}
Nancy Flournoy was born in Long Beach, California, on May 4, 1947.
After graduating from Polytechnic School in Pasadena in 1965, she
earned a B.S. (1969) and M.S. (1971) in biostatistics from UCLA.
Between her bachelors and masters degrees, she worked as a Statistician
I for Regional Medical Programs at UCLA. After receiving her master's
degree, she spend three years at the Southwest Laboratory for Education
Research and Development in Seal Beach, California. Flournoy joined the
Seattle team pioneering bone marrow transplantation in 1973. She moved
with the transplant team into the newly formed Fred Hutchinson Cancer
Research Center in 1975 as Director of Clinical Statistics, where she
supervised a group responsible for the design and analysis of about 80
simultaneous clinical trials. To support the Clinical Division, she
supervised the development of an interdisciplinary shared data software
system. She recruited Leonard B. Hearne to create this database
management system in 1975 (and married him in 1978). While at the
Cancer Center, she was also at the University of Washington, where she
received her doctorate in biomathematics in 1982. She became the first
female director of the program in statistics at the National Science
Foundation (NSF) in 1986. She received service awards from the NSF in
1988 and the National Institute of Statistical Science in 2006 for
facilitating interdisciplinary research. Flournoy joined the Department
of Mathematics and Statistics at American University in 1988. She moved
as department chair to the University of Missouri in 2002, where she
became Curators' Distinguished Professor in 2012.

While at the Cancer Center, Flournoy documented the
graft-versus-leukemia effect in humans and discovered a source of
frequent lethal viral infections in the bone marrow transplant
patients. Later she was influential in developing adaptive experimental
designs. Her numerous honors include fellow of the Institute of
Mathematical Statistics (1990), the American Statistical Association
(1992), the World Academy of Arts and Sciences (1992) and the American
Academy for the Advancement of Science (1993). She has received the
COPSS Scott (2000) and David (2007) awards, and the Norwood (2012)
award from the University of Alabama.
\end{abstract}

%
\end{frontmatter}

\section{Early Life}

{\bf Rosenberger:} Tell us a little about your early life. Where did
you grow up and what did your parents do?

{\bf Flournoy:} I was born in Long Beach, CA, and grew up in Los
Angeles County in a lemon orchard surrounded by oil wells and a flood
plain. There was a dairy farm nearby and we had a donkey. My father was
a plumbing contractor who plumbed Los Angeles: restaurants,
dormitories, cemeteries. He had 11 trucks go out every day. My Mom was
always unhappy about not finishing college, so she enrolled in college
when I went to college and then directed a preschool for many years. I
have three brothers and one sister. I was sent to Polytechnic School in
Pasadena as a sophomore in high school. On the entrance exam I had the
second highest score in math in history, but I flunked the English exam
because I didn't know the words in the instructions. (So even then I
had a one-sided brain!)

\begin{figure}

\includegraphics{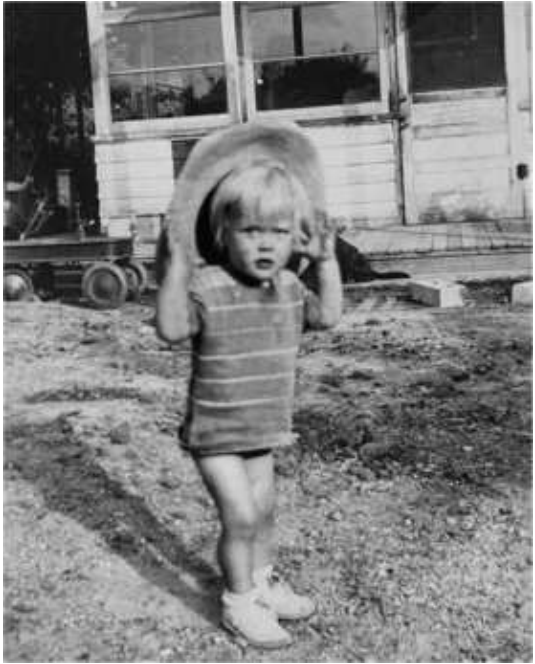}

\caption{Nancy at home in a part of Los Angeles County that was then
called Potero Heights, 1949.}\label{baby}
\end{figure}
\begin{figure}

\includegraphics{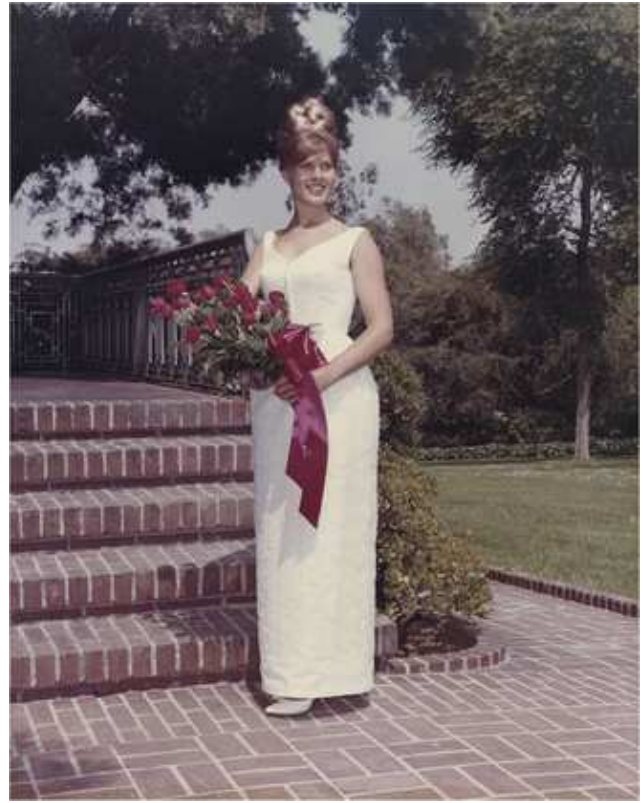}

\caption{Nancy at her graduation from Polytechnic School, Pasadena, 1965.}\label{hsgraduation}
\end{figure}

{\bf Rosenberger:} As a young person, were you interested in
mathematics, statistics, data? What made you excited about statistics?

{\bf Flournoy:} High school algebra really made me happy; I would lay
on the floor and work problems for hours. I had a new female instructor
whose husband had gotten a professorship across the street at Cal Tech
while she just landed a high school job; her anger came through and I
got the message that mathematics is worth being passionate about.

My love of statistics came as a junior at UCLA, when I took a course
taught by Don Ylvisaker. I just assumed that Don was a great teacher
for all time, but he later told me that he never had another class like
it. Four or five students from that class went on to get doctorates in
statistics.

\begin{figure*}[b]

\includegraphics{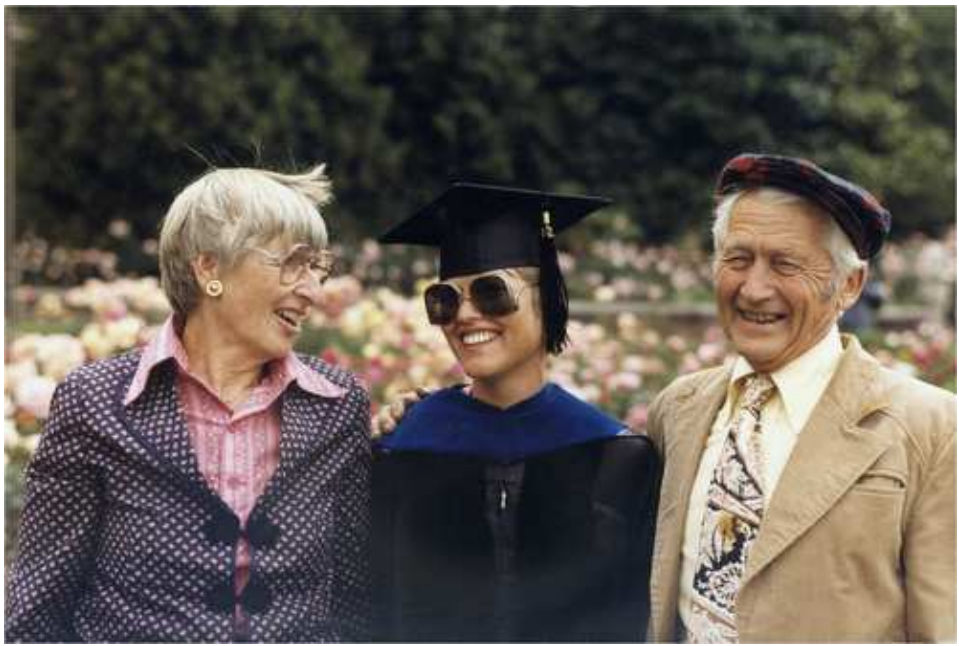}

\caption{Nancy with her parents, Elizabeth Blincoe and
Carr Irvine Flournoy, at her graduation from the University
of Washington in 1982.}\label{uwgraduation}
\end{figure*}

{\bf Rosenberger:} You were fortunate to be at UCLA at a time when
there were some of the great names in biostatistics: Abdelmonen Afifi,
Frank Massey, Wil Dixon, Olive Dunn, Virginia Clark. What professors
excited you at UCLA?

{\bf Flournoy:} Afifi was the young dynamic professor and taught out
of Scheff\'e; all the students loved Afifi. Dixon had a bimodal
distribution among the students; you either loved him or hated him. He
put out a thousand ideas a minute; if you paid close attention, you
would find they were pearls. It was a challenge to get what he was
saying as he didn't change the tone of his voice when he switched from
one topic to another. He taught the power of data analysis as a tool
for learning and a thousand little ways to make the data sing. I had a
class with Frank Massey; I learned a lot, but he was quiet and not dynamic.

{\bf Rosenberger:} Did you have any connection to the Department of
Statistics? You mentioned Ylvisaker. What about Paul Hoel?

{\bf Flournoy:} A separate statistics department did not exist at that
time; it was a math department with a few statisticians. I used the
Hoel, Port and Stone probability book when it was just a set of notes.
I don't think Hoel was the instructor though.

{\bf Rosenberger:} What interested you in biostatistics?

{\bf Flournoy:} Most of the statistics courses that were offered at
UCLA were in the Department of Biostatistics. Prior to taking
statistics, I had loved biochemistry and was a nutrition major, leading
to my major in the School of Public Health (SPA). When I recognized
that with a degree in nutrition, I would probably only be able to run a
cafeteria in a hospital, I decided to get my degree in mathematics
instead. I applied repeatedly to change from SPA to the College of Arts
and Sciences (CAS), but my application would get turned down. In tears,
I didn't know what to do. Then the SPA Dean asked why I was flunking
out, which didn't make sense since I always had gotten As and Bs. They
had lost all my records because I had changed names when I was
previously married, and nothing followed me. So that's why they didn't
accept me at CAS. By the time this got settled, I had enough credits to
get a degree in biostatistics.

\section{Graduate School}

{\bf Rosenberger:} What did you do after you graduated? How did you
get to graduate school?

{\bf Flournoy:} When I got a job at Regional Medical Programs as a
Statistician I, one old man would come around and ask me to add
numbers; I told him he could hire a statistical clerk for half my
salary. I was told that, as a young woman, my presentations were not
credible. So they hired a male DrPh to present my reports in his name.
As a mild way of protesting, I put my hair in a bun, dyed it white, and
got fired. They said I was an ``uppity woman.'' At that time, Virginia
Clark was department chair. She said, ``We have a fellowship, why don't
you come to grad school?'' I have some happy memories of my master's
program at UCLA: Olive Dunn supervised my master's thesis; Mary Ann
Hill was a great teaching assistant for Dixon's class; Carol Newton
taught a mean FORTRAN programming course; and Ray and Jean Mickey were
influential in my career decisions.

{\bf Rosenberger:} When you won the David Award, you talked about
meeting F. N. David. Tell us about that.

{\bf Flournoy:} I was in the Los Angeles chapter of the ASA; around
1972, a group of us carpooled out to UC Riverside where David was
giving a talk. She had a strong presence, standing with one leg up on a
stairstep and smoking a cigar while she talked. It was a roomful of
people, and she exuded such confidence. So I immediately started
smoking cigars. I had been used to seeing female statisticians such as
Clark and Dunn behind a desk and not commanding an audience.

{\bf Rosenberger:} How did you get to University of Washington (UW)?

{\bf Flournoy:} After the M.S., I thought I knew everything about
statistics. I got a job at Southwest Education and Laboratory for
Research, where there were a lot of education psychologists who were
into experimental design. On my second day, they presented me with
computer output that had more than one error term; I had the good sense
to keep my mouth shut. I immediately called Wil Dixon and asked what
they were talking about. He replied, ``Oh well, we can't teach you
everything.'' He suggested I get a book by Walt Federer. The book was
out of print, but Walt got a preprint from India and sent it to me; so
I spent my nights reading Federer's book.

Later, I was trying to read the {\em Journal of the American
Statistical Association} to implement some of the stuff I wanted to do,
and I found I couldn't read the literature. I also wanted to escape the
smog of Los Angeles. So I applied to the UW, my only application. Dick
Kronmal said there was a research assistant position with the bone
marrow transplant team, which was then located in the Old Public
Hospital (recently Amazon) in Seattle.

At that time, there was no statistics department at UW. The
mathematical statistics courses were taught in the Department of
Mathematics. I took the mathematical statistics sequence from Galen
Shorack. I had courses from Ron Pyke and Fritz Schultz in
nonparametrics. Shortly after Fritz left for Boeing, the remaining
statistics faculty formed the Department of Statistics. In the
Department of Biostatistics, there were some female faculty: Paula
Diehr and Pat Wahl. I took the first categorical data analysis class
taught at UW from Norman Breslow. He gave quizzes at the end of class,
so I never paid so much attention in a course before. I took survival
from Ross Prentice early in the days of the Cox proportional hazards model.

{\bf Rosenberger:} What was it like working with your dissertation
advisor, Lloyd Fisher?

{\bf Flournoy:} It worked out well because we have similar work
styles. Both of us had busy consulting lives; we would schedule
meetings and get our business done.

\section{The Seattle Bone Marrow Transplantation Team}

{\bf Rosenberger:} Today every street corner seems to have a contract
research organization for data coordinating centers on large clinical
trials. But when you went to the Fred Hutchinson Cancer Research
Center, information technology was primitive, such places did not
exist. You had to create that environment on your own. What was it
like? What were the challenges?

\begin{figure*}[b]

\includegraphics{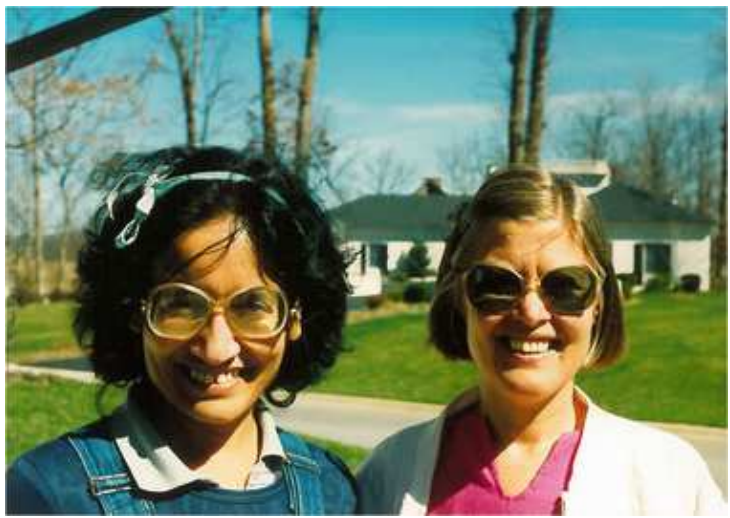}

\caption{Yash Mittal, first female director of the probability program,
and Nancy, first female director of the statistics program, at NSF.}\label{yash}
\end{figure*}

{\bf Flournoy:} That's an interesting story. Dick Kronmal had invested
a lot of effort in creating a database management system without
requiring a rectangular data structure. Updates required physically
sorting the cards (remember all data records had to fit into the 80
digits of a Hollerith punch card---so I tend to use the words ``card''
and ``record'' interchangeably). There was a transplant data set in
place with seven different kinds of cards. Kronmal had E. Donnal ``Don''
Thomas (Director of the Clinical Research Division at the Cancer
Center) buy a computer. The computer weighed 50 pounds (I could toss it
in my van and take it home; the cost was about \$50,000), and data
storage was on Phillips cassette tapes. Records could be transmitted
across the phone wires and then integrated into the database at UW.
Initially, there was not much data (only 10 patients) and the first
update took my whole computing budget for the year! What I did then for
some period of time was, when it was time to do an update, punch cards
of the whole database and the new dataset; I would physically sort and
merge the cards by hand and load them into SPSS. That was my ``dirty
laundry'' story because the laundromat had big long tables and I sorted
cards while doing laundry. Kronmal told me that, if I had any trouble
with my new computer, I should call Leonard Hearne. Index sequential
files were brand new at that time, and Leonard used them to create an
early database management system before the word was in the literature
(see Flournoy and Hearne, \citeyear{FloHea,FloHea90N1,FloHea90N2}). We used it for several
years until a commercial system came on the market. At site visits,
someone would ask a question and I would pass a note down to a
programmer, who would extract the answer in 15 minutes or so. We set
the bar for oncology programs.

Ross Prentice came from the University of Waterloo with a box of cards
on the Cox proportional hazards model; we were really early using that.
The doctors were smart enough to understand the limitations in using
discriminant analysis and they were thrilled to be able to incorporate
censored survival data in their regression models. My work documenting
graft vs. leukemia in humans was very important (see Weiden et al.,
\citeyear{Weietal79,Weietal81N1,Weietal,Weietal81N2}).
One hypothesis motivating the development of bone marrow
transplantation was that the marrow graft would attack residual
leukemia also. Immunological activity of the graft was apparent when
the graft instigated an immunological attack on the patient. I modeled
the impact of this attack on the relapse rate. The protection of the
graft attack against relapse greatly complicated post-transplant
treatment strategies. But our findings have withstood the test of time.
It was, perhaps, the first major application of the proportional
hazards model with time-dependent covariates.

{\bf Rosenberger:} When you think of the success of the bone marrow
transplant program (Don Thomas won the Nobel prize in 1990 for
developing bone marrow transplantation as a treatment for leukemia),
how much did statistics and data management play a role in that? Do you
think statistics and data management will ever get its due?

{\bf Flournoy:} We had this rudimentary set of records that could be
added onto infinitely. It started out that bacteriology wanted to add a
card, then virology, then specific studies would add a card with their
data. Before you knew it, we had an interdisciplinary shared database
with assigned patient numbers so all the integrated data was available.
I was able to say ``do you know what they're doing in virology that's
related'' because I knew everybody's data. It wasn't until many years
later that people started talking about having integrated shared
databases. Most were established for billing purposes, not for research
purposes. They are different constructs. Hospitals would archive data
after the bill was paid but we wanted to keep it around forever.

When the program started, there was one of everybody (one statistician,
one virologist, etc.), and we would sit around the table and share
results. It was important to be influential and to catch problems in
data collection and quality control before they got big. When working
with new doctors, there were humps you had to get over because they
would claim that there were no quality control issues: their lab people
never made a mistake. A lot of negotiation had to go on before we could
agree. Yes, we had a huge influence. Even randomization and blinding
was controversial. If it was in the middle of the night the cards might
get shuffled; there was too much room for bias. We introduced them to a
very careful randomization regimen for treatment assignments, with a 24
hour on-call person.

It will be hard for statistics and data management to ever get its full
due because the doctors are so enamored of themselves (laughs). It's
really a strange system where the people with the least science
background usually run the science. Also, the data management budget
was always the first to be cut; yet it is very expensive to do a
quality job.

{\bf Rosenberger:} What has your role been in fostering
interdisciplinary research?

{\bf Flournoy:} Having conducted interdisciplinary research for more
than a decade at the Cancer Center, I knew the power that teams of
interdisciplinary researchers could bring to bear on important
scientific questions. Coincidentally, when I went to the National
Science Foundation (NSF) in 1986, the Division of Mathematical Sciences
(DMS) had funded the Institute of Mathematical Statistics (IMS) to
write a report on cross-disciplinary research. I watched the growth in
their thinking as they interacted with each other. At the time, the
discipline did not appreciate the role of applications in academic
settings. I think I was able to influence the IMS cross disciplinary
committee on the valuable nature of interdisciplinary work. The report
of the committee had a dramatic effect on the discipline. The report
proposed establishing the National Institute of Statistical Science.
Since I was at NSF, I was able to promote the idea of establishing a
broad institute that would work on problems of national importance.

At the same time, I would receive proposals from statisticians
motivated by applications. Because our budget was small, I took such
proposals around to the relevant disciplines that were involved, and
was able to get some joint funding. This resulted in my getting an
award in 1988 for facilitating the funding of these interdisciplinary
projects. This also led to specific DMS requests for proposals for
interdisciplinary research projects, which are now common throughout NSF.

\section{Adaptive Designs}

{\bf Rosenberger:} How did you get interested in adaptive designs?

{\bf Flournoy:} While at the Cancer Center, the major program project
grant had five-year reviews. When we prepared for the third one of
these, we spent a year reviewing what we had done and how we would go
forward. In the course of that retrospective, I developed some feelings
about the two arm clinical trial. The standard ideas about the two arm
clinical trials came from the Peto paper in the mid 70s (Peto et al.,
\citeyear{Petetal76,Petetal77}). But, in my experience, a treatment is a point in a high
dimensional space: involving drugs, radiation, including how much, how
often; and one learned little about this high dimensional space using
the traditional two arm clinical trial. For instance, we spent five
years comparing A to B; but then we go back to the high dimensional
space and pick out point C, and have another five years of
experimentation and compare A to C. Then we compare C to D, and after
15 years we have knowledge of four points in a high dimensional space.
I believed it would be more efficient and informative to know which
direction we should head in the high dimensional space. So that led me
to think about adaptive designs. I recommended several to the group and
the physicians liked the ideas, but thought they may be too radical to
get funded.

Another thing that promoted my interest was looking at pilot studies to
decide what to take forward to larger studies. Bob Tsutukawa was
visiting the Cancer Center from the University of Missouri at the time.
I thought his Bayesian ideas were appealing and I used expert opinion
for prior elicitation (see Flournoy, \citeyear{Flo93}). The prior was way off, so
we wound up with a lot of toxicities. You just can't trust the best
expert opinion of the best experts, and so there needed to be some way
to use interim data faster to adapt and put much less weight on the
prior. My later work showed how random walk rules could be constructed
to do this (see Durham and Flournoy, \citeyear{DurFlo94}; Durham, Flournoy and
Rosenberger, \citeyear{DurFloRos97}; Flournoy and Oron, \citeyear{FloOro}).

\begin{figure*}

\includegraphics{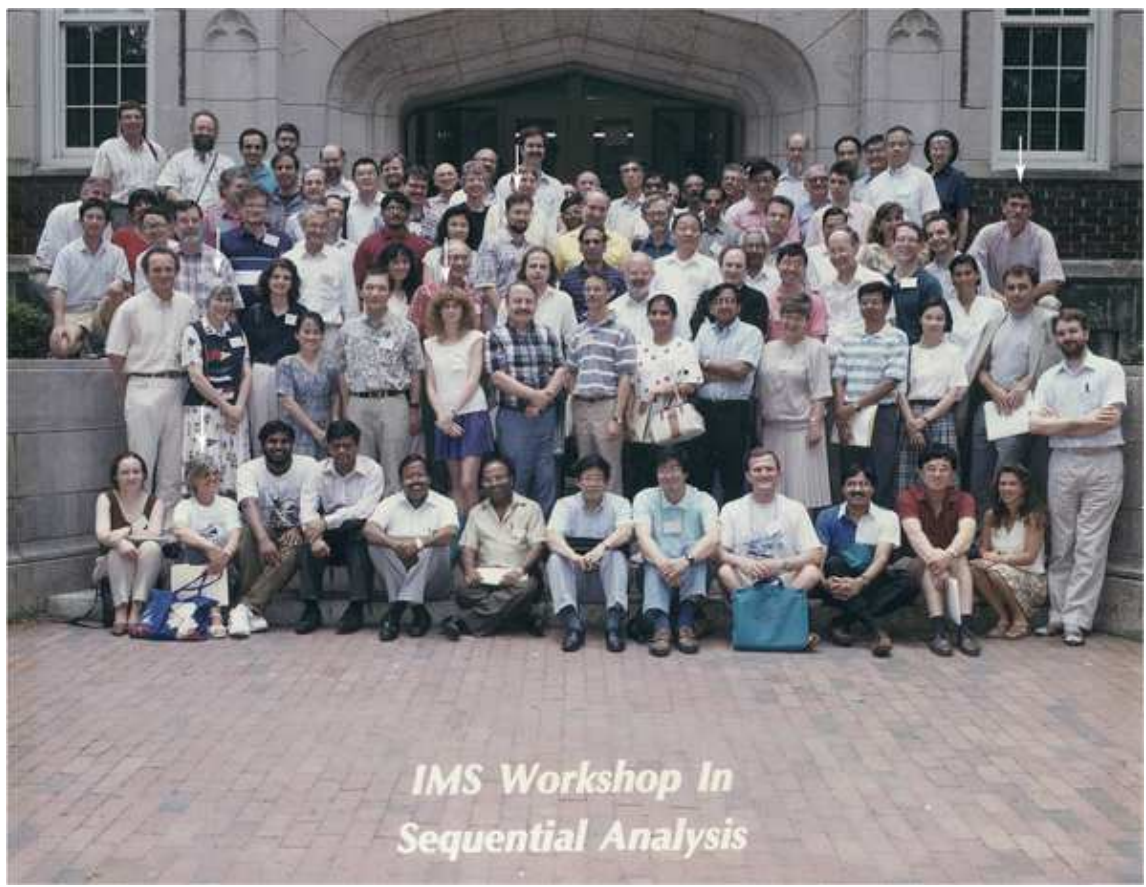}

\caption{1994 IMS Workshop on Sequential Analysis at University of North Carolina, Chapel Hill.  Included are Nancy Flournoy (seated second from left), Lynne Billard (immediately behind Nancy), Janis Hardwick (to the right of Lynne),
Bill Rosenberger (third row from back, middle), and Steve Durham
(directly in front of right window).}\label{seqanal94}
\end{figure*}

{\bf Rosenberger:} The first time I heard the name Nancy Flournoy was
in the context of the 1989 session on adaptive designs at Joint
Statistical Meetings (JSM) in Washington. It turned out to be one of
the most controversial sessions in the history of JSM. Talk about that.

{\bf Flournoy:} My experience at NSF was that you don't make progress
without community. One person alone doesn't get much done. So I had the
idea that a JSM session on adaptive design would bring together people
who are interested in adaptive designs. I didn't know anyone
personally. I invited based on my impressions of their interests. I
invited Don Berry, Richard Simon and Janis Hardwick. I gave a
straightforward technical talk on the topic. The remainder of the
session focused primarily on criticism of the extracorporeal membrane
oxygenation (ECMO) trials (e.g., Barlett et al., \citeyear{Baretal85};
O'Rourke et al.,
\citeyear{ORoetal89}; Ware, \citeyear{War89}).
(The ECMO trial was an implementation of the
randomized play-the-winner rule of Wei and Durham, \citeyear{WeiDur78}, in which 11
babies were assigned to an experimental arm, and all survived, while
one baby assigned to the conventional arm, died. The historical death
rate on the conventional arm was 80 percent.) To my dismay, all the
negative focus of the session was directed toward the adaptive design
aspect of the clinical trial, rather than on the sample size and what
kind of sample size would be needed for the trial to be convincing. The
press that was generated by this session set adaptive designs back a
long time.

{\bf Rosenberger:} How much do you think the failed ECMO trial
inhibited the development of adaptive designs?

{\bf Flournoy:} What would have been a reasonable approach? The
original trial was unconvincing due to having few patients, in spite of
the fact that a probabilistically reasonable stopping rule was applied.
The controversy over the subsequent two arm trial in clinical research
set back adaptive designs wrongly. The adaptive trial was so successful
that only one baby died; is that bad?

{\bf Rosenberger:} In your 1992 AMS/IMS/SIAM conference on adaptive
designs (Flournoy and Rosenberger, \citeyear{FloRos1995}), you brought together some of
the  groundbreakers of adaptive designs along with a number of younger
faculty who are now at the forefront of the discipline. At the opening
session, you started by talking about the need to streamline the
process of clinical trials, the end to phases and the incorporation of
dynamic interim decisions. You said that will revolutionize the way we
do clinical trials, and that this conference would be an ambitious
beginning to that revolution. Now, over two decades later, there are
70-some sessions on adaptive designs at the Joint Statistical Meetings,
``big-pharma'' working groups, Food and Drug Administration white papers
and guidelines, companies like ADDPLAN, and CYTEL devoted to adaptive
design software and everyone wants to do adaptive designs. What took so long?

{\bf Flournoy:} ECMO made a steep hole to climb. We also had to
develop theory. It was one thing to say ``this is a good idea,'' and
another to adequately support it. Some ideas were NOT good. This
includes a class of procedures that derive from stochastic
approximation, that Val Fedorov coined ``best intention'' designs. In
these designs, a target dose is estimated (such as the dose having a
particular percent toxicity or one that maximizes some utility
function); then that estimate is the dose given to the next subject.
Some, including Lai and Robbins (\citeyear{LaiRob82}), understood early on that using
this procedure without safeguards may result in treatment sequences
that converge to the wrong dose. But others, including myself (Li,
Durham and Flournoy, \citeyear{LiDurFlo95}), were enamored of this idea and ignorant of
earlier warnings. This approach remains popular today even as recent
publications are exposing just how misleading it can be (e.g., Azriel,
\citeyear{Azr12}, Oron and Hoff, \citeyear{OroHof13}).

In the 1980s, John Whitehead spent a year visiting the Cancer Center
from the University of Reading, and promoted the idea of using
sequential stopping rules taking censoring into account. Its value was
so obvious that I expected that by 1990 every clinical trial would be
using these techniques. So I focused instead on adaptive allocation. At
American University (AU), I worked on theoretical problems in these
areas. When I pulled my head out and looked around I was shocked to see
that stopping rules incorporating censoring were not being used, except
a bit in cancer. So things that seem obvious to some can take a long
time to enter the medical arena. Take the ``$3+3$'' dose escalation design
as an example. It has been soundly discredited (Reiner, Paoletti and
O'Quigley, \citeyear{ReiPaoOQu99}; Lin and Shih, \citeyear{LinShi01}),
and yet remains a standard
practice in oncology phase I trials.

Adaptive allocation is still in its infancy compared to sequential
monitoring and stopping. Now there has developed a new belief that
simulation is adequate for assessing an adaptive design. But relying
solely on simulation muddies the water because there is no global view
of what is driving the design. In addition, there are many papers in
the literature that report only averages over simulations without
measures of variability. When you consider measures of variability, a
completely different picture emerges (Oron and Hoff, \citeyear{OroHof13}). I fervently
believe in developing the theory underlying classes of designs.
Fortunately, many people are interested in working on the theoretical
challenges, and there are a lot of interesting open questions.

{\bf Rosenberger:} Many times when I hear talks on adaptive designs I
want to scream out ``Nancy Flournoy thought of that in the 1980s.'' How
do you feel about some of your early ideas being ignored?

{\bf Flournoy:} Well, I'm hardly alone in this. For instance, Chris
Jennison invented many clever techniques for sequential and adaptive
clinical trials very early that are sometimes ``rediscovered'' without
reference (e.g., Jennison, Johnstone and Turnbull, \citeyear{JenJohTur82};
Kulkarni and
Jennison, \citeyear{KulJen86}; Jennison, \citeyear{Jen87}).
In my case, it amazes me that there
are a large number of people who will reference a paper from the 1980s
and ignore 30 years of my research. For example, the early up and down
paper of Storer (\citeyear{Sto89}) is often cited without reference to my later
papers that have much more sophisticated control of the adaptive
process. This early paper is used as a whipping post to declare up and
down procedures inferior. An up and down design is a random walk that
can end anywhere. The last state (dose) visited should not be used as
an estimator. But this is done when the up and down design is compared
to other procedures that derive from stochastic approximation (e.g.,
Zacks, \citeyear{Zac09}). That bothers me a lot.

{\bf Rosenberger:} How did you meet Steve Durham? This began one of
the great collaborations in statistics. Tell us about that.

{\bf Flournoy:} One of the few positive consequences of the 1989 JSM
session was meeting Steve Durham from the University of South Carolina.
When I walked out the door after the session, Steve introduced himself
and was very excited because we were basically working on the same
mathematical problems, his from an engineering motivation, and mine
from a medical motivation. We began working together right away. He
would come to Washington, DC, to meet me, and I went to South Carolina.
After a stint as Chair at AU, I was on a sabbatical at the University
of North Carolina Chapel Hill; Leonard and I bought a house close to
campus so that we could host visitors. In particular, Steve Durham and
I worked together quite a lot in that house and at the Department of
Statistics. Several other collaborators came down for extended periods,
including you (W.F.R.) and two of my doctoral students: Eloi Kpamagen
(now at Novavax) and Misrak Gezmu (now at National Institutes of Health).

{\bf Rosenberger:} The introduction of the random walk rules coincided
with the introduction of the continual reassessment method (CRM;
O'Quigley, Pepe, and Fisher, \citeyear{OQuPepFis90}) in the Bayesian context. In
particular, you and Steve worked out the entire exact distribution
theory of a class of designs, while others were relying on simulation.
How does this rank in terms of your contributions to statistics?

{\bf Flournoy:} The random walk rules are extremely practical and
mathematically elegant, so it was a lot of fun to develop the theory.
They are the standard in many areas of science (e.g., American Society
for Testing and Materials, \citeyear{ASTM1991}; Treutwein, \citeyear{Tre95};
National Institute of
Environmental Health Sciences, \citeyear{NIEH2001}).
The key property that we
discovered was how to control the allocation coverage by introducing an
appropriate bias (Durham and Flournoy, \citeyear{DurFlo94}). Steve was always
thinking in terms of engineering applications; I was always thinking in
the dose-response context. We did ``reverse engineering,'' in that we
had a target allocation in mind, and we found design parameters to
facilitate this. The designs are nonparametric in that allocation does
not depend on estimates of model parameters. They are extraordinarily
simple to illustrate and have exact distribution theory that is
unavailable for other, more complicated designs.

{\bf Rosenberger:} Some have lumped random walk rules in the context
of generic dose escalation designs, such as the $3+3$ design, that has no
optimal properties. At the same time, Bayesian approaches, such as the
CRM were becoming increasingly well-known. Talk about the historic
interplay among these approaches.

{\bf Flournoy:} Lloyd Fisher and John O'Quigley (from the University
of Leeds) were hired at the Cancer Center to replace me when I left for
the NSF. Lloyd and I laughed that it is not often one's dissertation
advisor replaces his student! John was initially responsible for
implementing a random walk rule that I had designed in a pilot study
for a bone marrow clinical trial. He let them get away with a simple
dose escalation procedure, but he and Lloyd got introduced to the
subject at that time. They immediately thought of doing a Bayesian
alternative, and it was published in 1990 in {\em Biometrics}
(O'Quigley, Pepe and Fisher, \citeyear{OQuPepFis90}); the major random
walk paper
appeared in {\em Biometrics} in 1997 (Durham, Flournoy and
Rosenberger, \citeyear{DurFloRos97}). Most of the Bayesian literature
 was, by necessity,
simulation based, whereas Steve and I were busy obtaining a complete
workable probabilistic theory of the random walk procedures.

There are a number of philosophical differences among the approaches.
Fedorov would call the CRM a ``best intention'' approach, because it
involves predicting a target dose and treating the next patient at that
dose, sequentially. Our approach is estimation-motivated. The idea is
to get allocations into a region of interest that allows us to
efficiently estimate the dose-response curve in that region.

There is also a short-memory and long-memory distinction: allocation
probabilities for the random walk rule converge exponentially fast to
their asymptotic limits. Alternatively with best intention designs
(which to date are long-memory designs), nonrepresentative early
allocations can cause the design to converge to the wrong dose (see,
e.g., Azriel, Mandel and Rinott, \citeyear{AzrManRin11};
Oron, Azriel and Hoff, \citeyear{OroAzrHof11};
Azriel, \citeyear{Azr12}). Such phenomena were observed early on in the context of
stochastic approximation designs (e.g., Lai and Robbins, \citeyear{LaiRob82};
Bozin
and Zarrop, \citeyear{BozZar91}).

Adaptive optimal designs are promising long memory designs, but they
depend on parameter estimates to get started. Random walk procedures
that target optimal design points provide good start-up information
with small sample sizes. Alternatively, one can regularize the
information matrix, a ``fix'' that is often called ``Bayesian designs''
even though no posterior distribution is obtained. True Bayesian
estimator updates coupled with dose allocations made in some stable
optimal way, rather than in a ``best intention'' way are also promising.

{\bf Rosenberger:} What is the future of adaptive designs? Do you
think all clinical trials will eventually be adaptive?

{\bf Flournoy:} I think there is a great future for adaptive designs.
I think experimentation will always involve a series of trials; the
question is how well one utilizes information from one to the next.
There is a lot of value in relatively small but sequential trials (see
Flournoy, \citeyear{Flo}), because these trials involve many design features,
including the grid size and range on which you are operating. The best
use of one experiment may be to tell you how you could have better
selected design characteristics; then you can refine the estimate of
the target of interest.

\begin{figure*}[b]

\includegraphics{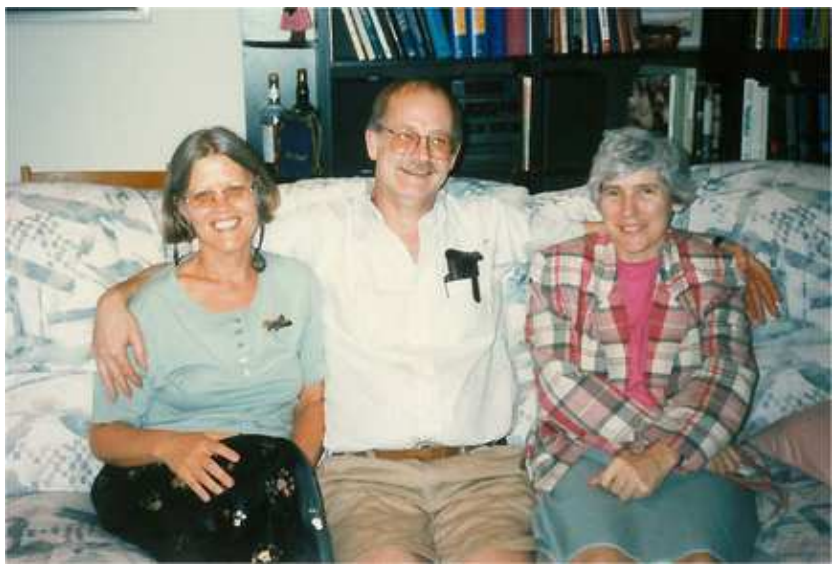}

\caption{Nancy, husband Leonard, and Lynne Billard at their home
in Chapel Hill, NC, 1994.}\label{lenandlynne}
\end{figure*}

Some of my work has been on inference and estimation following adaptive
designs (e.g., Rosenberger, Flournoy and Durham, \citeyear{RosFloDur97};
Ivanova and
Flournoy, \citeyear{IvaFlo}; May and Flournoy, \citeyear{MayFlo09};
Lane, Yao and Flournoy, \citeyear{LYF2014}).
One has to be careful doing everything sequentially because some of the
interim changes may cause final estimates to lack normality. For
example, in best-intention designs, the estimate of a slope parameter
can march off to infinity for some common models. Also, even if an
adaptive dose-finding procedure has a fixed total sample size, the
sample sizes at each dose are random variables. In up-and-down
procedures, the proportion of subjects allocated to each dose tends to
a constant and standard asymptotic normality results. But in many other
adaptive designs, proportions of subjects allocated to each dose tend
to a random variable. This causes the conditional information matrix to
be random, even in the limit, in which case standard conditions for
asymptotic normality fail. These are many interesting questions to be
explored about adaptive designs.

\section{Women in Statistics}

{\bf Rosenberger:} Talk about the creation of Pathways to the Future,
its successes, and its legacy.

\begin{figure*}

\includegraphics{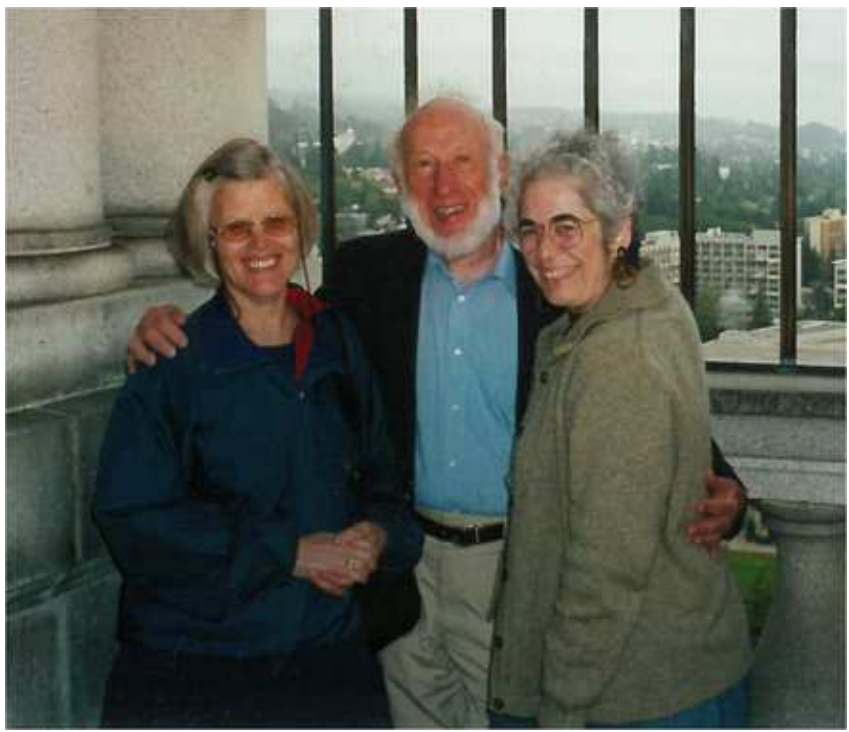}

\caption{Nancy, Ingram Olkin and Elizabeth Margosches
(formerly with the Environmental Protection Agency) at the
Campanile at University of California, Berkeley, 2003.}\label{ingram}
\end{figure*}

{\bf Flournoy:} I went to the 1984 Annual IMS Meeting in Lake Tahoe.
At that meeting, there were five women out of about 200 attendees. It
became quite clear to me that this was an important place for academic
statisticians to meet and focus on academic interests. In anticipation
of the 1988 Fort Collins IMS meeting, which was separate from the Joint
Statistical Meetings, I decided it would be great to see more women
there. So I bounced ideas off Mary Ellen Bock (Purdue University) and
Lynne Billard (University of Georgia). Lynne agreed to take the lead in
organizing a workshop for women at the upcoming IMS meeting. Lynne had
the brilliant idea of having Elizabeth Scott (University of California,
Berkeley) give the keynote lecture. At that time, we were debating
whether there was any gender inequity in academia, and we weren't sure.
I had never experienced problems at UCLA or UW. However, when I went to
the NSF, Yash Mittal (the first female director of the probability
program) and I saw that there were almost no female grantees, and very
few were even applying for grants.

The evening presentation by Scott really hit us very hard: she had tons
of data and randomized studies on gender inequity. Any questions about
inequities in how women were recruited, judged and valued were thrown
out the window. Scott's way of handling this lecture was wonderful
because she went through all this horribly depressing data, but she
then turned around and finished the lecture by telling us what we could
do to protect ourselves. She ended with two positive notes: that
outcomes are not predetermined, and one can take one's career in one's
own hands. Lynne ran the workshop for the next two decades, and she
presented Scott's lecture with updated data every year. That lecture
was the last lecture Scott gave before she passed away. I remember well
that there was a palpable sigh of relief from Scott---that she could
turn over her cause to the next generation.

{\bf Rosenberger:} How did you become NSF program director? What was
your experience with gender issues there?

{\bf Flournoy:} Ingram Olkin has long been a great friend and mentor.
He is the one who recommended me to the NSF for the program director
position. I was the first female director in the statistics program the
same year that Yash Mittal was the first female probability director.
Some people had indicated to the division director their fear I was
going to give all the grant money to biostatistics. I convinced him
that I could represent the entire statistics field.

One day I remember answering the phone and a professor on the line
yelled ``I said I wanted to speak to the director,'' thinking a woman
on the phone must be a secretary.

We had a good travel budget and I went to as many young women's
lectures as I could. I would go up at the end of their talk and ask if
they would be interested in applying for a grant. By the time I left
NSF, the proportion of grant proposals from women was proportional to
their presence in the field. A suggestion is such a small thing, and
yet clearly important messages weren't being transmitted to female faculty.

{\bf Rosenberger:} Was discrimination subtle or not so subtle when
your career was developing?

{\bf Flournoy:} Well, there was always sexist behavior and many things
that were said and done are considered inappropriate or even sexual
misconduct today. When I went on the job market for a fully academic
position I found that many men were incredulous. Some would make
outrageous comments directly to me as if I were invisible (or a man).
Men in my own age category were often dismissive or oblivious to my
presence. Some of the older generation was very helpful and supportive
(I think of Shanti Gupta, Purdue University; Norman Johnson, University
of North Carolina at Chapel Hill; Lucien LeCam, University of
California, Berkeley; Ingram Olkin, Stanford University; and Manny
Parzen, Texas A\&M University). The younger generation just thought of
me as another senior person, so they were fine.

{\bf Rosenberger:} What is your feeling about the role of women in
statistics today? I can say, from my perspective on 20 years of search
committees, that from a hiring perspective, we are thrilled to have
qualified women candidates and compete hard to get them. And certainly
policies on tenure to allow maternity leave have vastly improved over
the years, as have the composition of committees and senior
administrators. Is there any work left to be done?

{\bf Flournoy:} You can see improvement, but there are still troubling
facts: just try to find a woman in the 2013 JSM awards brochure, for
instance. Women are getting hired at proportional rates now, but
awards, tenure and advancement are areas where there much is left to be
done. See Lynne Billard's new update of Scott's old data on the subject
(Billard and Kafadar, \citeyear{BilKaf14}). That will depress you.

\section{Conclusion}

{\bf Rosenberger:} You talked a little about your transition into a
fully academic position. The latter part of your career was spent at AU
and University of Missouri (MU), and considerable time as department
chair, and a mentor to many diverse
students. Talk about this.

{\bf Flournoy:} AU was a great place for me when I went there in 1988.
I had left the Cancer Center with a staff of 23, a budget of \$700,000
and responsibilities that had become a burden when I became convinced
of the need for more nimble learning strategies in dose-finding
clinical trials. I had eight doctoral students at AU, and all but two
of them developed mechanisms to control random walks and urn models,
and to provide mathematical descriptions of their controlled behavior.
One worked on issues of inference following an adaptive design and one
worked on a problem in economics. I am proud that four of these
students are black and two are women.

\begin{figure}

\includegraphics{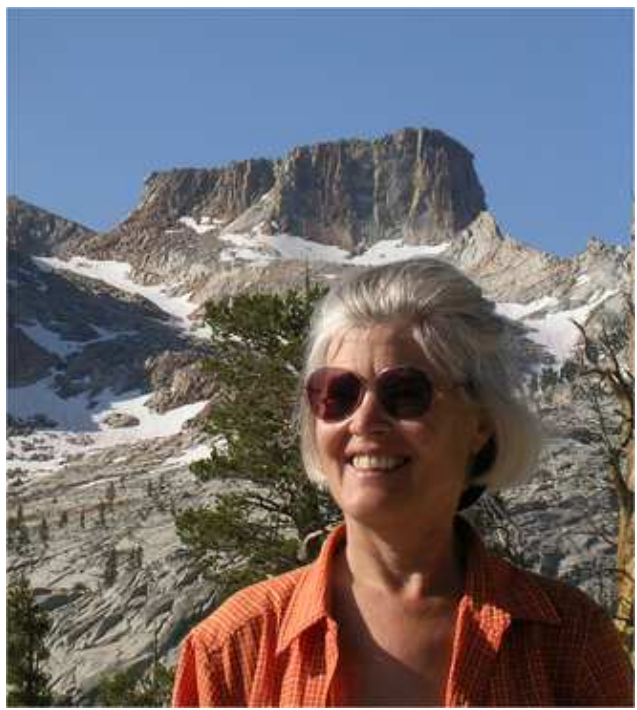}

\caption{Nancy near Aasgard Pass in the Enchantment Lakes Wilderness
Area, Washington, where she was hiking with her husband Leonard and her
colleague Lori L. Thombs (University of Missouri) following the 2006
Joint Statistical Meetings in Seattle.}\label{hiking}
\end{figure}

Unfortunately, a very destructive president came to AU, and by 2000 it
was clear that STEM graduate programs were going to be dismantled. AU
had one of the oldest statistics doctoral programs in the country and
it was sad to see it threatened by ignorance and arrogance. To remain
in a department with a doctoral program, I needed to move and this led
me to accept the chair at Missouri in 2002. When I stepped down as
chair in 2011, I had doubled the number of tenure-track faculty and
added five teaching faculty positions. I increased the presence of the
department across campus through joint appointments and a targeted
increase in service courses, and I increased the prestige of the
department nationally, personally promoting our faculty and enabling
their participation in national and international activities. More
details can be found in a Chapter I recently wrote on the history of
statistics at MU (Flournoy and Galen, \citeyear{FloGal}).

I have graduated seven doctoral students from MU. We worked on adaptive
and optimal designs; we developed new models for specific, challenging
dose-response problems and we have illuminated the effect of having
dose allocations depend on the history of prior allocations and
responses. My students continue to bring me a great deal of pleasure.

{\bf Rosenberger:} What are your hobbies and interests?

{\bf Flournoy:} I love hiking. I am not happy with a trip that takes
less than four days. A four-day trip has two days out and two days
back---so one is never very far from a road. After hiking for more than
two days, one must rely on one's self much more completely. It is so
peaceful. I gave up trying to hike in the East and the Midwest United
States. One just can't get far enough away from roads; and the
mountains aren't high enough. I like trekking around timberline for a
week or more where the views are spectacular. I keep going back to
Yosemite, Kings Canyon and Sequoia National Forests. Nepal was great,
too. I try to get in one long hike each year. In the meantime, I~dance.
I~resumed ballet classes while at AU; it is great mind-to-body exercise
and wonderful for strength and balance. Leonard and I enjoy English
country dance together. Throw in Pilates and yoga and I am happy.

To survive a severe health challenge that had the doctors stumped, I
gained considerable knowledge of alternative methods and became
accomplished in some. But that is another story.

{\bf Rosenberger:} What's next for Nancy Flournoy?

{\bf Flournoy:} Well I have a lot of ideas. I'm really interested in
questions of inference following adaptive designs. We have some
examples in two stage designs that maximum likelihood estimators are
mixtures of normals; some designs lead to estimators that are normal
with random variances. I think our preliminary results are
generalizable, but this remains to be shown. I'm optimistic that
tractable solutions to seemingly intractable problems are at hand.



%

\end{document}